\documentclass{pazhastl}

\usepackage{mathptmx}
\usepackage{graphicx}

\def\smfigurewocap#1#2#3{
  \begin{minipage}{0.049\linewidth}
    \rotatebox{90}{#3}
  \end{minipage}
  \begin{minipage}{0.95\linewidth}
    \includegraphics[bb=45 185 560 688,width=0.97\linewidth]{#1}
    \centerline{#2}
  \end{minipage}
}

\def\smfigurewocapp#1#2#3{
  \begin{minipage}{0.049\linewidth}
    \rotatebox{90}{#3}
  \end{minipage}
  \begin{minipage}{0.95\linewidth}
    \includegraphics[bb=45 185 560 510,width=0.97\linewidth]{#1}
    \centerline{#2}
  \end{minipage}
}

\begin{document}

\journalinfo{2003}{29}{9}{573}[578]

\title{First hours of the GRB 030329 optical afterglow}

\author {R.~A.~Burenin\email{burenin@hea.iki.rssi.ru}\address{1},
  R.~A.~Sunyaev\address{1}, 
  M.~N.~Pavlinsky\address{1}, 
  D.~V.~Denissenko\address{1}, 
  O.~V.~Terekhov\address{1},
  A.~Yu.~Tkachenko\address{1},
  Z.~Aslan\address{2},
  I.~Khamitov\address{2},
  K.Uluc\address{2},
  M.~A.~Alpar\address{3},
  U.~Kiziloglu\address{4},
  A.~Baykal\address{4},
  I.~Bikmaev\address{5},
  N.~Sakhibullin\address{5},
  V.~Suleymanov\address{5}
  \addresstext{1}{Space Research Institute (IKI), Moscow, Russia}
  \addresstext{2}{T\"{U}BITAK National Observatory, Antalya, Turkey}
  \addresstext{3}{Sabanci University, Istanbul, Turkey}
  \addresstext{4}{METU, Ankara, Turkey}
  \addresstext{5}{Kazan State University (KGU), Kazan, Russia}
}
\shortauthor{BURENIN \etal}
\shorttitle{GRB 030329 OPTICAL AFTERGLOW}
\submitted{28 April 2003}

\begin{abstract}

  We present the first results of the observations of the extremely bright
  optical afterglow of gamma-ray burst (GRB) 030329 with the 1.5m
  Russian-Turkish telescope RTT150 (T\"{U}BITAK National Observatory,
  Bakyrlytepe, Turkey). RTT150 was one of the first 1.5m-class telescopes
  pointed to the afterglow. Observations were started approximately 6 hours
  after the burst. During the first 5 hours of our observations the
  afterglow faded exactly as a power law with index $-1.19\pm0.01$ in each
  of the BVRI Bessel filters. After that, in all BVRI filters
  simultaneously we observe a steepening of the power law light curve. The
  power law decay index smoothly approaches the value $\approx-1.9$,
  observed by other observatories later. This power law break occurs at
  $t-t_0\approx0.57$~days and lasts for $\approx \pm0.1$~days. We observe no
  variability above the gradual fading with the upper limits $10$--$1\%$ on
  time scales $0.1$--$1000$~s. Spectral flux distribution in four BVRI
  filters corresponds to the power law spectrum with spectral index
  $\alpha=0.66\pm0.01$. The change of the power law decay index in the end
  of our observations can be interpreted as a signature of collimated
  ultrarelativistic jet. The afterglow flux distribution in radio, optical
  and x-rays is consistent with synchrotron spectrum. We continue our
  observations of this unique object with RTT150.


\end{abstract}

\section*{Introduction}
\label{sec:intro}

Although the main part of the energy of gamma-ray bursts is emitted in hard
X-rays and gamma, optical observations allowed to obtain a very important
information on their sources. For example, they allowed to measure the
redshifts for a number of GRBs establishing the cosmological distance scale
to the sources of gamma-ray bursts. The observations of the GRB optical
afterglows have become possible only when rapid and accurate GRB
localizations were obtained in X-rays (first with \hbox{{\em Beppo}SAX}
satellite, e.g. Costa \etal\ 1997a). The first optical afterglow was
discovered by Bond (1997). The GRB afterglows were found in other
wavelengths from radio to gamma as well (e.g. Taylor \etal\ 1998, Costa
\etal\ 1997b, Burenin \etal\ 1999).

A number of the optical afterglows were observed in detail to date. Usually
the light curves at all wavelengths can be approximated as power laws or
their combinations with breaks at different times after the bursts. These
breaks are attributed to signatures of collimated ultrarelativistic jets in
GRB sources (see, e.g. references in Frail \etal\ 2001). Other type of the
afterglow variability is also observed and interpreted as signatures of
underlying supernovae (e.g. Sokolov 2001).

\begin{figure*}
  \vskip 5mm
  \begin{center}
    \includegraphics[height=0.45\linewidth]
    {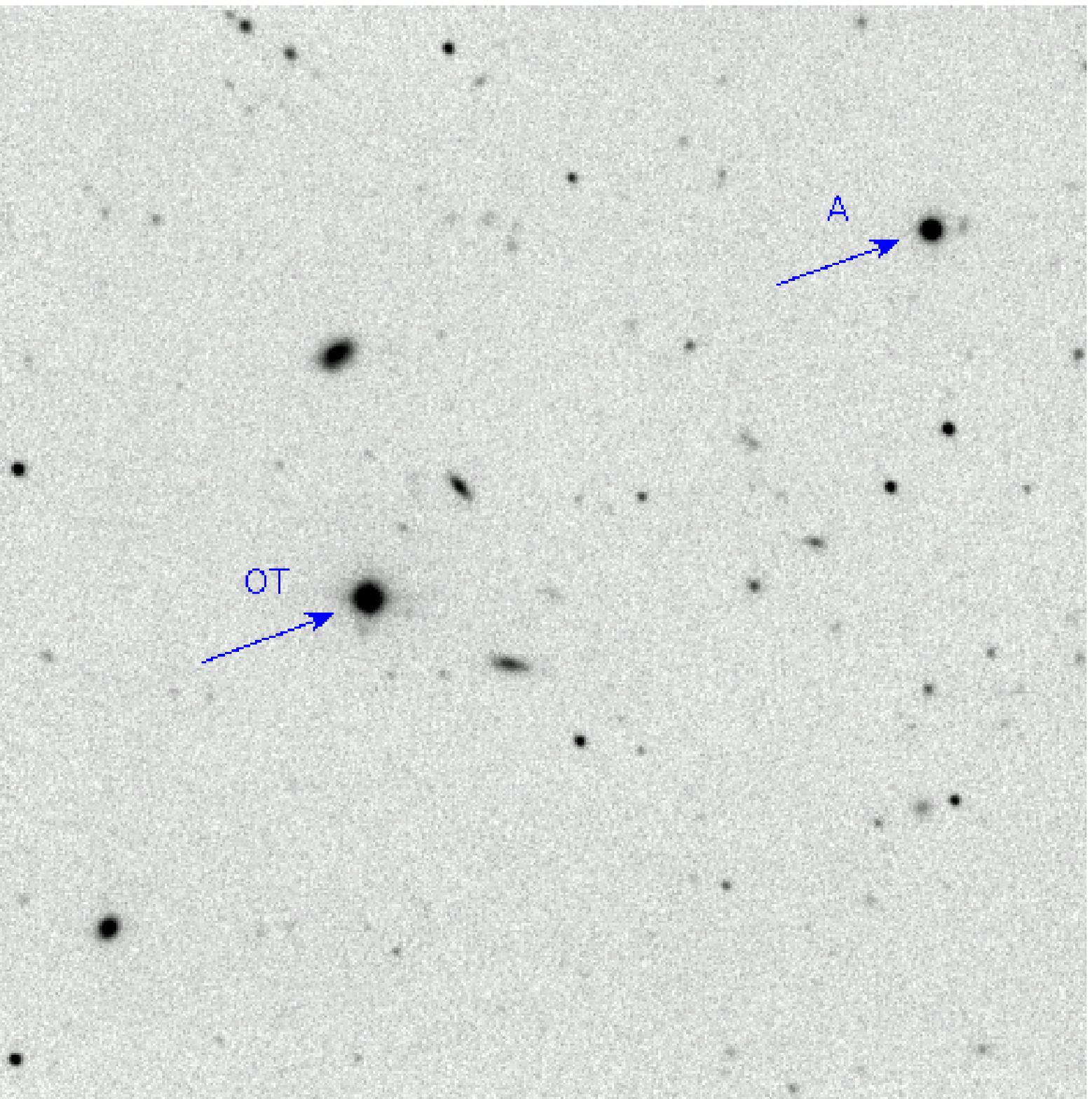}
    ~~~~
    \includegraphics[height=0.45\linewidth]
    {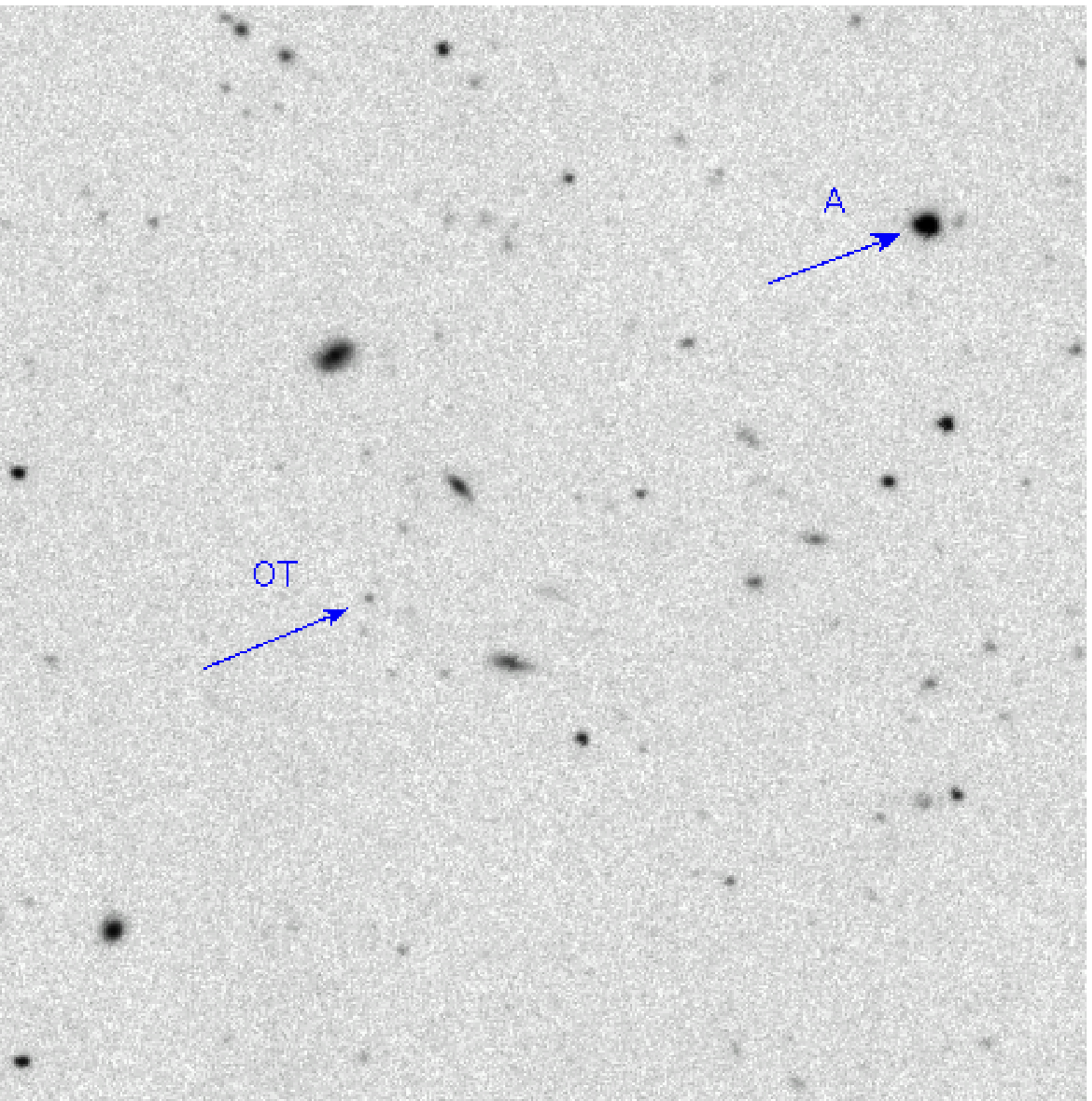}
  \end{center}
  \caption{The image of the GRB 030329 afterglow, obtained with RTT150
    telescope on March 29 (left) and 
    on May 28 (right). Arrows show the optical transient (OT) and the
    reference star (A) used to calibrate the flux of the OT (see
    text).\label{fig:fchart}}
\end{figure*}

A very bright gamma-ray burst 030329 was detected by instruments aboard
HETE-II satellite at 11:37:14.7~UT. The burst peak flux was measured to be
$7\cdot10^{-6}$~erg~s$^{-1}$\,cm$^{-2}$ in 30--400~keV energy range, and
duration --- about 30~s (Vanderspek \etal\ 2003, Golenetskii \etal\ 2003).
Within 2 hours a very bright optical transient (OT) was discovered in the
position consistent with that measured by HETE-II (Peterson \& Price 2003,
Torii 2003). At the moment this object has become a prime target for many
telescopes throughout the world. The extensive photometric and spectroscopic
observations of this object are well documented in GCN circulars
(e.g. Burenin \etal\ 2003, Khamitov \etal\ 2003). The redshift of the OT was
determined to be $z=0.1685$ (Greiner \etal\ 2003), ranging it as a closest
GRB with measured redshift (except for GRB~980425, if it is associated with
supernova 1998bw at $z=0.0085$). This distance to the source corresponds to
``isotropic'' energy release in gamma-rays of order $10^{52}$\,erg. The
underlying supernova component emerged in OT spectrum later (Stanek \etal\
2003).

In this paper we present the first results of the observations of this
extremely bright optical afterglow with the 1.5m Russian-Turkish telescope
(T\"{U}BITAK National Observatory, Bakyrlytepe, Turkey, 2547\,m,
$2^\mathrm{h}01^\mathrm{m}20^\mathrm{s}$~E, $36^\circ49'30''$~N).
RTT150 turns out to be one of the first 1.5m-class telescopes pointed to the
afterglow. Observations were started already in approximately 6 hours after
the burst. At the beginning of our observations the magnitude of the optical
afterglow in R was estimated to be $14^m$. This is $3$--$4^m$ brighter than
any GRB afterglow previously observed on this time scale.

\section*{Observations}
\label{sec:obs}

We used an AP47p Apogee CCD mounted in the Cassegrain focus of the telescope
($1:7.7$). This is a back-illuminated $1056\times1024$ CCD chip. We used
$2\times2$ binning mode with $0.46''$ pixels. As always, bias and dark
counts were subtracted and flat field correction applied for all the
images. The data reduction was done with IRAF (Image Reduction and Analysis
Facility)\footnote{http://tucana.tuc.noao.edu/} and using our own software.

\begin{figure*}
  \begin{center}
    \begin{minipage}{0.48\linewidth}
      \smfigurewocap{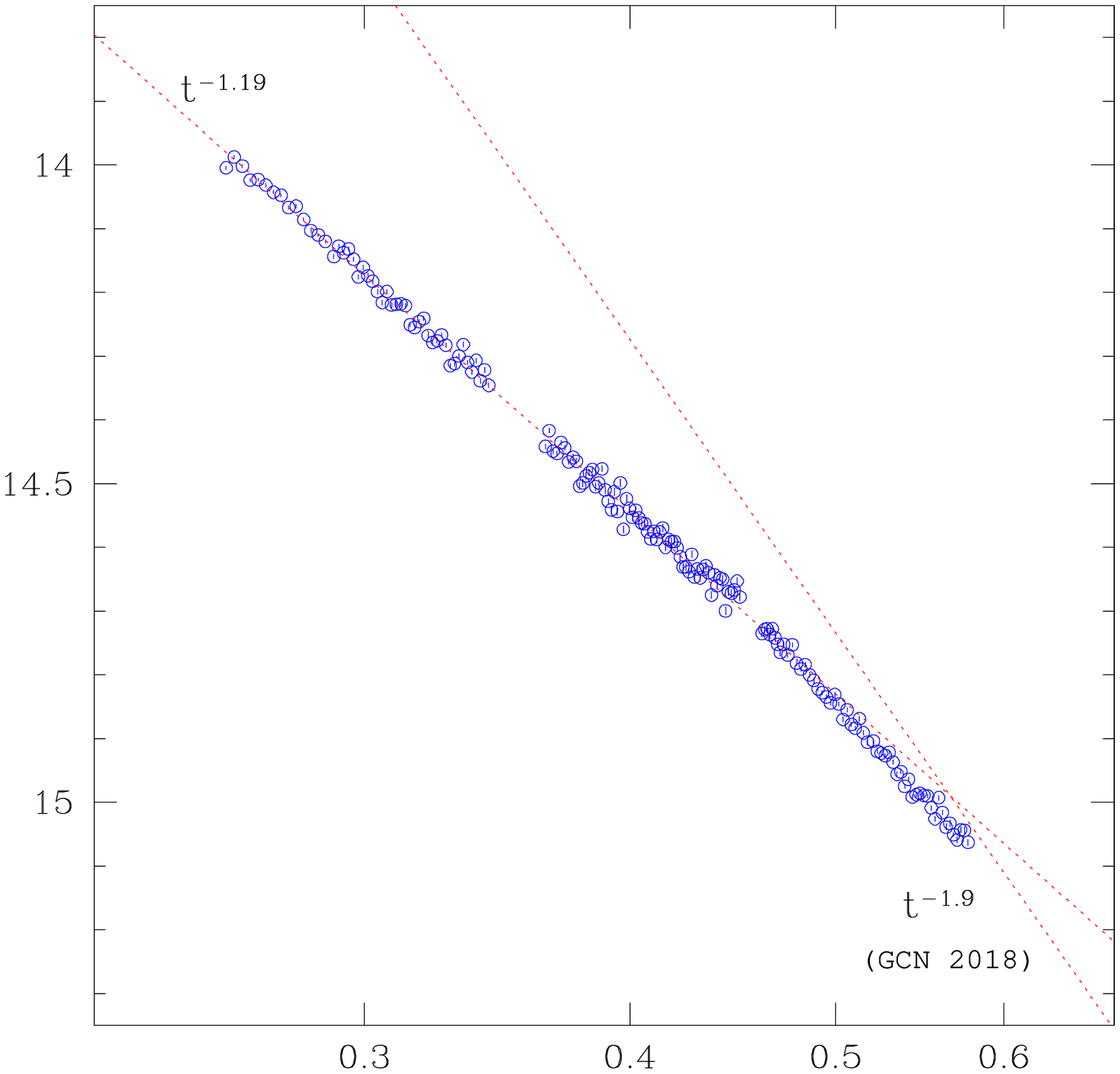}{$t-t_0$, days}{$m_R$}
    \end{minipage}
    ~~~~
    \begin{minipage}{0.48\linewidth}
      \smfigurewocap{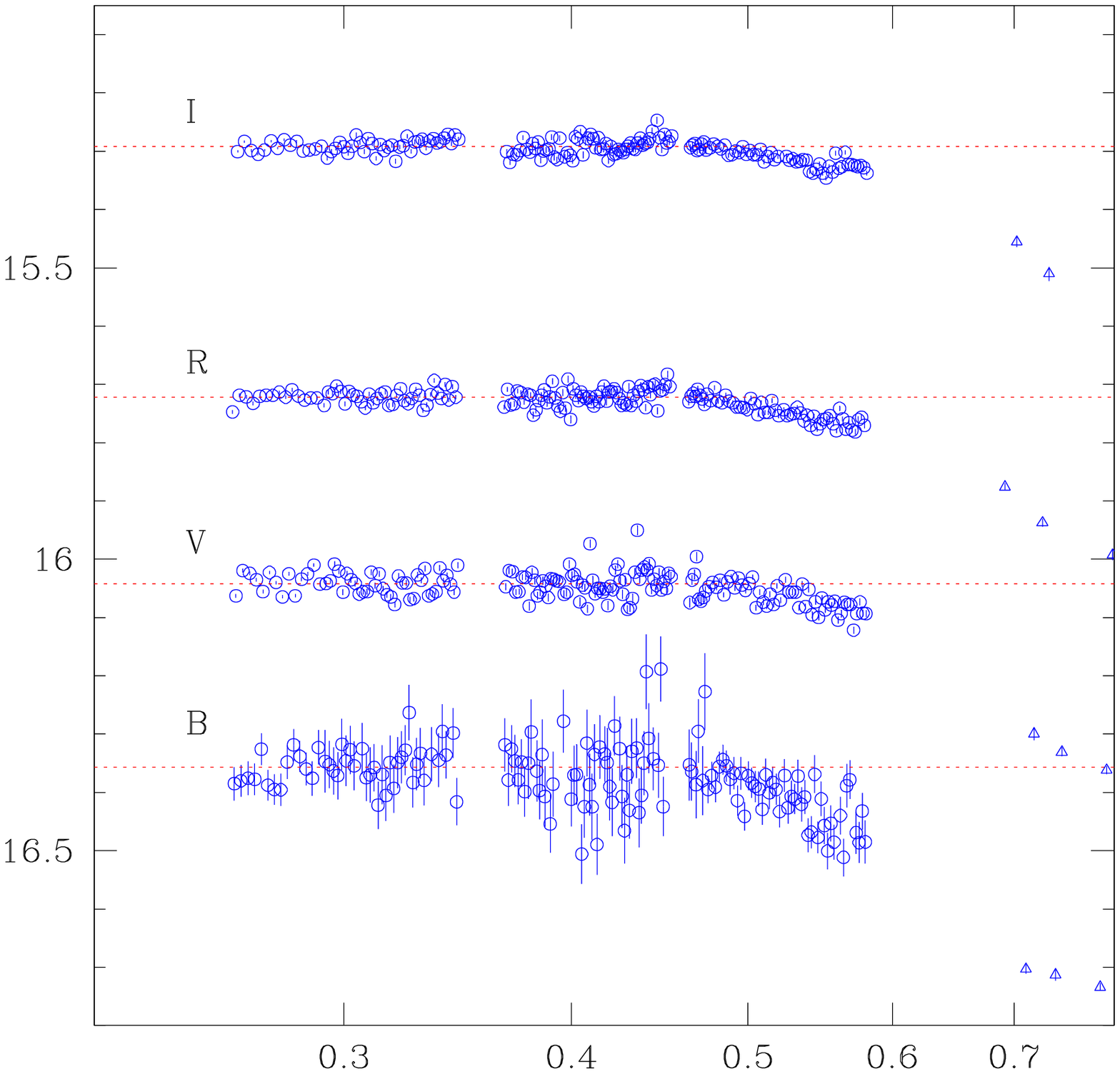}{$t-t_0$, days}{$m-2.934\,\lg\,t$}
    \end{minipage}
  \end{center}
  \caption{ The light curve of the GRB 030329 optical afterglow during the
    first night of the observations. The left panel shows the light curve in
    R band. The right panel gives the light curve in the BVRI bands after
    the subtraction of the power law decay with index $-1.19\pm0.01$,
    obtained by fitting the first 5~hours of the R band light curve. The
    left panel also shows the power law with index $-1.9$, obtained at
    $t-t_0>0.6$~days (Garnavich \etal\ 2003). In the right panel the
    triangles at $t-t_0>0.7$~days show the data from Fitzgerald \& Orosz
    (2003). \label{fig:lcl}}
\end{figure*}

At the very beginning of the night on 29 March 2003, at 17:50~UT, the
telescope was pointed to the afterglow of gamma-ray burst 030329. We
observed the object till 01:30~UT when its zenith distance had become higher
than $68^\circ$. All this time the OT field was imaged in BVRI Bessel
filters in series of exposures 30 -- 10~s (the readout time was about
10~s). In total about 700 images were taken, 175 in each BVRI filter. The
combined R image of the field is shown in Fig.~\ref{fig:fchart} in the
left. The right panel of this figure shows the image of this field obtained
on May 28. 

\section*{Photometry}
\label{sec:phot}

All the measurements of the OT flux during the first night of our
observations were done relative to the nearby bright star, designated as
``A'' in Fig.~\ref{fig:fchart}. This is the brightest star in our field of
view. Nevertheless, during the first night even the star ``A'' is much
fainter than the OT. Thus the errors of the afterglow flux measurements
during the first night are dominated by the flux errors for this star. All
the results were checked using the other stars in the field, but all of them
are much fainter than star ``A'' and can not be used to improve the OT
photometry.

To calibrate the field we observed the Landolt (1992) stars before and after
the afterglow observations. These observations show that night was almost
perfectly photometric. Small decrease of atmosphere transparency was
observed during the very beginning of the night. The transparency was lower
by approximately 10\% and returned to its usual value during the first few
hours of the observations. 

Our photometric calibration is in good agreement with the photometry of this
field provided by Henden (2003). In this circular it was noted that the OT
and star ``A'' have very different colors. Therefore using star ``A'' as a
reference may cause an additional systematic error correlated with airmass
due to possible differential airmass corrections. However, in our
photometric solutions we found no variations of the color coefficients with
airmass or with the decrease of the transparency during the first hours of
our observations. We estimate the systematic errors of the OT flux
measurements to be not more than 1\%.

The light curve of the GRB 030329 afterglow in R band, obtained in the first
night of our observations, is shown in the left panel of
Fig.~\ref{fig:lcl}. During the first 5 hours the afterglow flux decayed with
good accuracy as a power law with index $-1.19\pm0.01$. After that for
approximately 3 hours we observe a steepening of the power law light curve.
The power law decay index smoothly approaches the value $\approx-1.9$,
observed later by the observatories in the western hemisphere. The left
panel also shows the power law with index $-1.9$, obtained at
$t-t_0>0.6$~days with FLWO 1.2m telescope (Garnavich \etal\ 2003).

The right panel of Fig.~\ref{fig:lcl} shows the light curve in BVRI filters
after the subtraction of power law decay with index $-1.19\pm0.01$, obtained
by fitting the first 5~hours of the R band light curve. We see that the
power law decay is the same in each of the BVRI filters. The power law
slopes in BVI are equal to $-1.22\pm0.05$, $-1.19\pm0.02$ and $-1.19\pm0.01$
respectively, which is within the errors equal to the power law slope in
R. At $t-t_0\approx0.5$~days the flux decay in all filters simultaneously
starts to deviate from this power law. Closer to the end of our observations
we observe somewhat higher deviation in B. In the right panel of
Fig.~\ref{fig:lcl} the triangles at $t-t_0>0.7$~days show the data from
Fitzgerald \& Orosz (2003). These data are approximately consistent with our
measurements at $t-t_0<0.6$~days.

Comparing our data at $t-t_0<0.6$ with the data of western observatories at
$t-t_0>0.6$ (see Fig.~\ref{fig:lcl}) one can see, that we were lucky to
observe in detail the beginning of the power law break of the afterglow
light curve. The break occurred at $t-t_0\approx0.57$~days and lasted for
approximately $\pm0.1$~days. Note that just before the break we observe some
marginally significant flattening of the afterglow light curve in R and I.

Before our observations, at $t-t_0<0.25$, there were unfiltered observations
with small telescopes (e.g. Rykoff \etal\ 2003, Sato \etal\ 2003). These
data should be calibrated properly to be compared with our observations. We
note however that the power law slope of the afterglow light curve at
$t-t_0$ from 0.06 till 0.17~days (from 1.5 till 4 hours) equals to
$1.07\pm0.08$ as can be inferred from the ROTSE data (Rykoff \etal\ 2003),
which is approximately consistent with our measurements at later $t-t_0$. An
approximately similar slope was observed also in infrared (Nishihara \etal\
2003). Thus, it is likely that the the power law decay index of the
afterglow does not change significantly from as early as 1--2~hours after
the burst.

\begin{figure}
  \smfigurewocapp{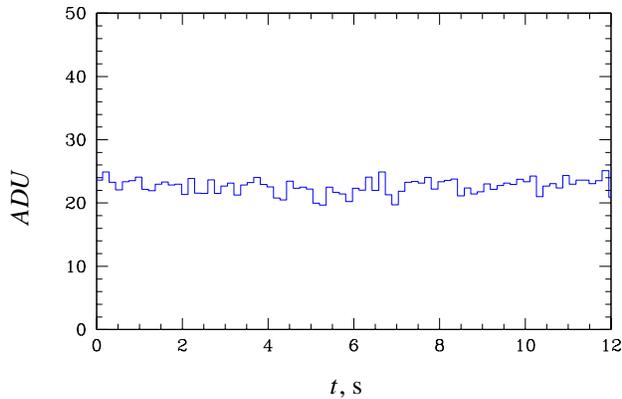}{$t$, s}{$ADU$}
  \caption{A part of the afterglow light curve, obtained with high temporal
    resolution.\label{fig:tr}}
\end{figure}

\section*{Short time scale variability}
\label{sec:var}

In Fig.~\ref{fig:lcl} one can see that we observe no variability of GRB
030329 afterglow flux above the gradual decay on time scales 100--1000~s.
Detailed analysis of the aperiodic variablity of the afterglow light curve
is beyond the scope of this paper and will be a subject of the forthcoming
publication. We note however, that RMS deviation of the observed light curve
in R band from the best fit power law equals to $\approx1\%$ on these time
scales. This number obviously includes contribution of the statistical and
systematical errors and therefore gives a conservative estimate of the upper
limit on the intrinsic variablity of the optical flux.

To examine the variability of the afterglow flux on even shorter time scales
we observed the OT field, turning off the hour tracking of the
telescope. The OT was moving through the field of view during these
exposures. We then reconstructed the OT light curve with temporal resolution
up to $0.1$~s using the track of the OT in CCD frame. In this method one
image contains a segment of the OT light curve of approximately 15~s
duration. Twenty images were taken in this way between 20:00 and
20:30~UT. One of those OT light curve segments is shown in
Fig.~\ref{fig:tr}. From these data we estimate the RMS of the afterglow
optical flux on $0.1$--$10$~s time scales to be equal to $10$--$3\%$
respectively, which again gives conservative upper limits on the intrinsic
variablity of the afterglow.

\begin{figure}
  \smfigurewocap{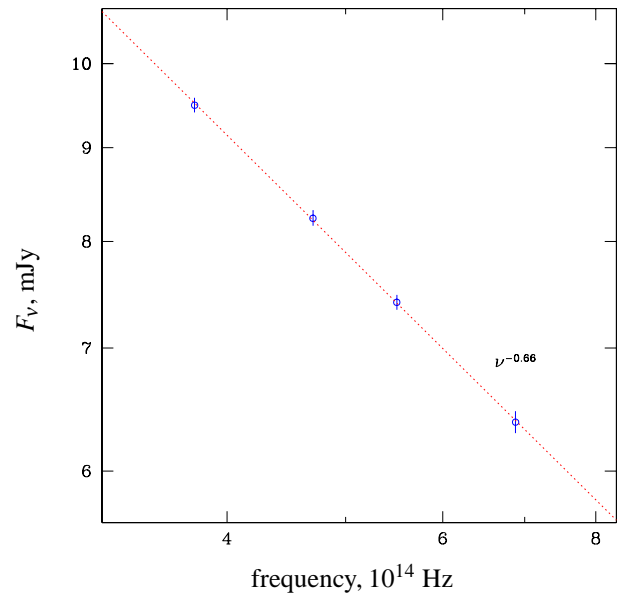}{frequency, $10^{14}$ Hz}{$F_\nu$, mJy}
  \caption{Spectrum of the afterglow at $t-t_0=0.25$~days
    (6~hours), inferred from the fluxes in four BVRI bands, corrected for
    galactic extinction $E(B-V)=0.025$ (Schlegel \etal\ 1998). This form of
    the spectrum remains exactly constant during the first 5~hours of our
    observations.\label{fig:sp}}
\end{figure}

\section*{Spectrum Across the BVRI Bands}
\label{sec:spectrum}

In Fig.~\ref{fig:lcl} one can see that the colors of the optical afterglow
of GRB 030329 are constant within the errors during the first 5 hours of our
observations. The low resolution spectrum obtained from the fluxes in four
BVRI filters at $t-t_0=0.25$~days, and corrected for galactic extinction
$E(B-V)=0.025$ (Schlegel \etal\ 1998) is shown in Fig.~\ref{fig:sp}. With
good accuracy it can be described as a power law with spectral index
$\alpha=0.66\pm0.01$. We observe a somewhat flatter spectrum than that
obtained by Stanek \etal\ (2003), who measured a power law spectral index
$\alpha=0.85$, 2.6~days after the burst. Note that both indices are not
corrected for the extinction in GRB host galaxy.

\begin{figure} 
  \smfigurewocap{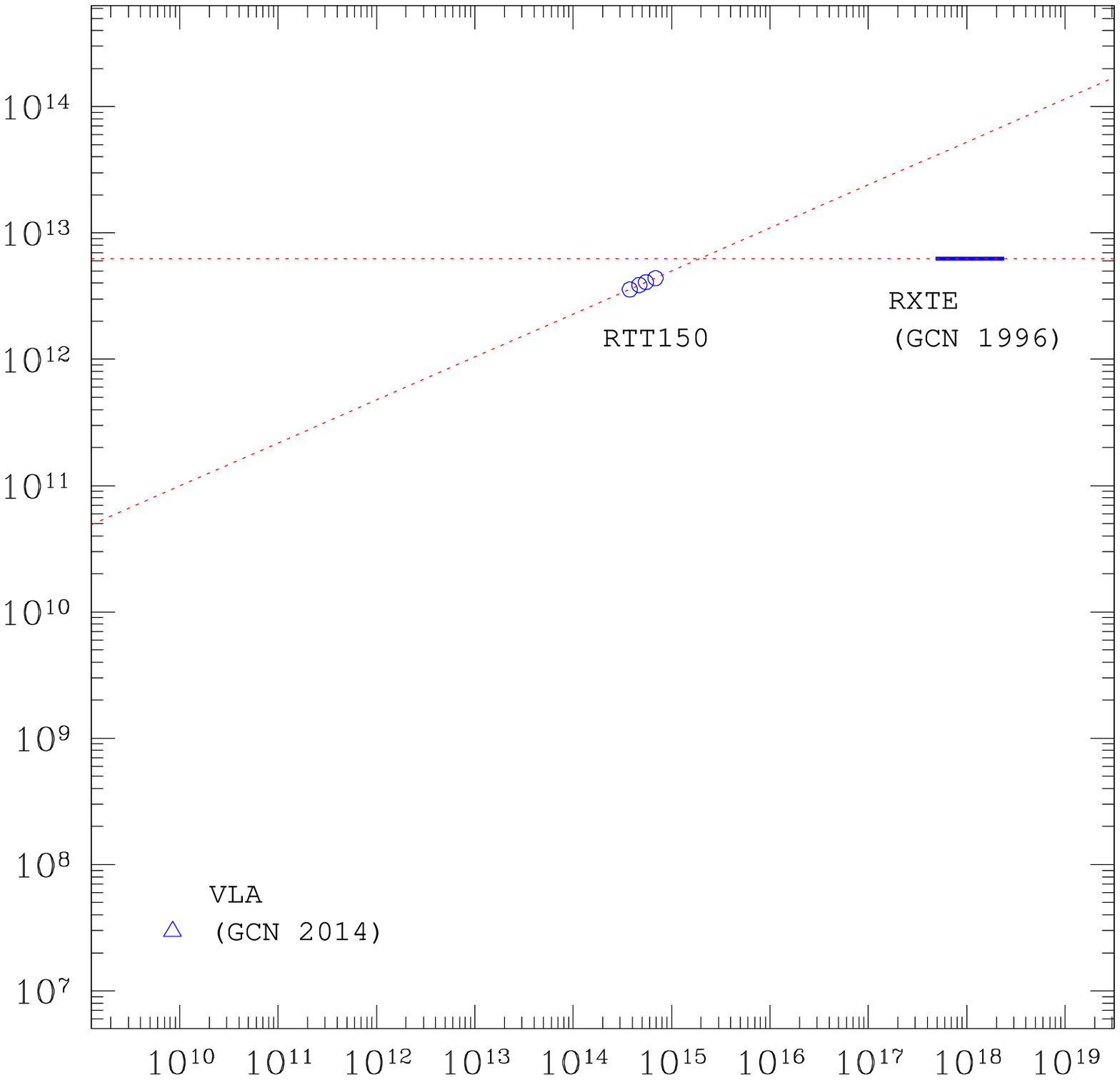}{frequency, Hz}{$\nu F_\nu$, Jy\,Hz}
  \caption{Afterglow $\nu F_\nu$ spectrum in radio, optical and
    X-ray bands. The RTT150 and RXTE data (Marshall \& Swank 2003) are shown
    as was observed at $t-t_0=0.25$~days (6~hours). The radio data
    correspond to $t-t_0=0.6$~days (14~hours) were taken from Berger \etal\
    (2003).
    \label{fig:spall}}
\end{figure}

In Fig.~\ref{fig:spall} we compare the afterglow fluxes in the optical, in
X-rays and in radio. The RTT150 data are shown for $t-t_0=0.25$~days
(6~hours), when the second RXTE observation was made (Marshall \& Swank
2003). We take the X-ray flux and the power law spectral slope from this
circular. According to this circular between $t-t_0=5$ and $6$~hours the
X-ray flux decayed by 20\%. This corresponds to a power law with slope
approximately $-1$, which is close to what we observe in the optical at
about the same time.

From Fig.~\ref{fig:spall} we see that at $t-t_0=0.25$~days the maximum of
the afterglow power is approximately in X-rays. The maximum is very wide
covering far ultraviolet and probably soft gamma-rays. Even in the optical
the afterglow power is only a factor of two lower than the power in X-rays.

The first observation of this afterglow in the radio was made approximately
0.6~days (14~hours) after the burst. The afterglow flux at 8.46\,GHz was
measured to be 3.5\,mJy (Berger \etal\ 2003). Between $t-t_0=0.25$ and
0.6~days the afterglow radio flux probably had not changed by more than an
order of magnitude. The afterglow radio flux was observed to be raising
after the first observation (e.g. Pooley 2003). Therefore, at $t-t_0=0.25$
the afterglow power in radio should be approximately by 5--6 orders of
magnitude lower than in optical or X-rays.

\section*{Discussion}
\label{sec:discussion}

We present a high sensitivity observation of the light curve of the GRB
030329 optical afterglow, starting as early as 6~hours after the burst. In
each of the BVRI filters we observe a gradual flux decay, which can be
accurately described as a power law $F_\nu\propto t^{-1.19}$ during the
first 5~hours of our observations. After that the afterglow flux started to
decline faster.

High signal-to-noise ratio during our observations allowed us to investigate
the variability of the afterglow flux with much higher temporal resolution
than it could be done previously. Within the errors of flux measurements we
observe no variability above the gradual fading down to the 0.1~s time
scale. The upper limits are 10, 3 and 1\% on 0.1, 10 and 100--1000~s time
scales respectively.

The results of our observations are consistent with the model where the
afterglow emission is generated during the deceleration of the
ultrarelativistic collimated jet (see, e.g. Hurley \etal\ 2003). The break
in power law light curve, which we observe at $t-t_0=0.57$, can be
interpreted as the ``jet break'', i.e. the break which occurs when the
angular structure of the ultrarelativistic collimated jet becomes
observable. This interpretation is supported by the fact that to good
accuracy this break occurred simultaneously in different colors. The power
law slope of the light curve changes from $-1.19$ to $-1.9$, approximately
as it is usually observed in jet breaks (e.g. Harrison \etal\ 1999).
Furthermore, before jet break there can not be any short time scale
variability --- exactly what we observe with good accuracy. The variability
indeed occurs after this break (see GCN circulars).

For a uniform jet moving toward the observer, the time of the jet break
corresponds to the time when the gamma-factor of the jet falls below
$\theta^{-1}$, the inverse opening angle of the jet. We can determine the
opening angle $\theta=0.08$ using the formula and typical parameters from
Frail \etal\ (2001). The actual energy release in gamma-rays appears to be
$10^{52}\theta^2/2=3\!\cdot\!\!10^{49}$~erg, approximately an order of
magnitude lower than the typical value obtained by Frail \etal\ (2001) and
comparable to the energy emitted by a typical supernova.

The form of the afterglow flux distribution in the radio, optical and X-ray
bands (Fig.~\ref{fig:spall}) is approximately consistent with synchrotron
emission. Here the synchrotron cooling frequency could be between optical
and X-rays. Synchrotron self-absorption is probably effective in radio
band.

The observations of this unique object are continued now with RTT150. The
results of these observations will be presented in the subsequent papers.

\bigskip

We are grateful to Mikhail Revnivtsev for his assistance in the analysis of
RXTE data and to Sergey Sazonov for useful discussion of the results of our
observations. This work was supported by Russian Fund for Basic Researches
(grants 02-02-16619, 03-02-06768, 02-02-17342), by Russian Government
Program of Leading Science Schools Support (grant
2083.2003.2), by the program of Presidium of Russian Academy of Sciences
``Variable phenomena in astronomy''. This work was also supported by the
High Energy Astrophysics Working Group of the Scientific and Technical
Research Council of Turkey (T\"{U}BITAK) through its support of basic
research in Turkish universities and by the Turkish Academy of Sciences (for
MAA).

\end{document}